\documentclass[11pt,a4paper]{article}
\usepackage[utf8]{inputenc}
\usepackage{amssymb}
\usepackage{graphicx}
\usepackage{subfig}
\usepackage{tabularx}
\usepackage[cmex10]{amsmath}
\usepackage{amsfonts}
\usepackage{array}
\usepackage{dblfloatfix}
\usepackage{balance}
\usepackage{flushend}
\usepackage{xcolor}

\usepackage{setspace}
\usepackage{bm}
\begin{document}

\title{Noise Resilient Recovery Algorithm for Compressed Sensing}
\author{\begin{tabular}{c c c}
V. Meena & and & G. Abhilash\\
meena\_pec10@nitc.ac.in & & abhilash@nitc.ac.in
\end{tabular} \\ \\
Department of Electronics and Communication Engineering \\
National Institute of Technology Calicut \\
Kerala, 673 601 India}

\date{}
\maketitle
\doublespacing
\begin{abstract}
  In this article, we discuss a novel greedy  algorithm for the recovery of compressive sampled signals under noisy conditions. Most of the greedy recovery algorithms proposed in the literature require sparsity of the signal to be known or they estimate sparsity, for a known representation basis, from the number of measurements. These algorithms recover signals when noise level is significantly low. We propose Entropy minimization based Matching Pursuit (EMP) which has the capability to reject noise even when noise level is comparable to that of signal level. The proposed algorithm can cater to compressible signals and signals for which sparsity is not known in advance. Simulation study of the proposed scheme shows improved robustness to white Gaussian noise in comparison with the conventional greedy recovery algorithms.  
\end{abstract}
\vspace{0.5cm}
\paragraph*{Keywords:}
Sparse representation, measurement matrix, entropy based matching pursuit, greedy recovery algorithms, compressed sensing 

\section{Introduction}
Compressed sensing (CS) is a signal acquisition scheme that embeds the intelligence of compression along with signal acquisition in discrete form. The utility of CS is more pronounced in acquiring wide-band signals which have a sparse representation in some domain. 
\par 
If $\{\bm\Psi_i(t);\;i=1,2,...,N\}$ is a representation basis for $\bm{S}(t) \in \mathbb{R}^N$, then $\bm{S}(t)= \sum_{i=1}^N{C_i\bm{\psi}_i(t)}$, where $C_i$, $i$ = $1,2,...,N$ are representation coefficients, which form an $N\times1$ vector $\bm{C}$ = $[C_1,C_2,...,C_N]^T$. For signals in $\mathbb{R}^N$, the $\ell_0$ norm is defined as 
\begin {equation}
\|\bm{C}\|_0 = |supp(\bm{C})| = |\{i:C_i \neq 0\}|. 
\end{equation}
$\bm{S}(t)$ is $K$-sparse if $\|\bm{C}\|_0 \leq K$ with $K\ll N$. Real signals are rarely sparse, but are compressible. A signal is compressible if it can be represented in an appropriate basis with only a few significant coefficients, \itshape{i.e}; \upshape when sorted in the descending order, the coefficients follow power-law decay as,
\begin{equation}
|C_i| \leq P i^{-r}, \;for\; i = 1,2,3,...\; and \;r > 1,
\end{equation}
where the non-negative constant $P$ is independent of $r$. The lower bound on the number of measurements  required for stable recovery is a function of the sparsity $K$ of the signal. The basic requirement in CS is to identify a representation basis relative to which sparsity $K$ is minimum. \\
A measurement matrix which is incoherent or least correlated with $\bm{\Psi}$ measures the signal at a rate much lower than the Nyquist rate required for the signal. If the number of measurements is greater than its sparsity, it would be possible to recover the signal in a stable manner \cite{CSC, bara}. To ensure that the geometry of sparse signals is preserved in the measurements, the matrix  $\bm{\Phi}$  of measurement functions should satisfy the Restricted Isometry Property (RIP) \cite{candtao}\cite{CSC}: 
\begin{equation}
 (1-\delta_K)\left\|\bm{S}(t)\right\|^2\leq \left\|\bm{\Phi} \bm{S}(t)\right\|^2 \leq 
 (1+\delta_K)\left\|\bm{S}(t)\right\|^2;\ 0 < \delta_K < 1,
 \label{orig_rip}
  \end{equation}
where $\delta_K$ is the restricted isometry constant (RIC) corresponding to the $K$ sparse signal $\bm{S}(t)$. The RIP requires the energy of the measurement to remain within a closed limit around the energy of the signal. The recovery of the sparse set of representation coefficients becomes more stable as $\delta_K$ approaches zero. The measurement vector $\bm{y}$ = $\left[y_1, y_2, ...., y_M \right]$ can be obtained as the inner products of the signal $\bm{S}(t) \in \mathbb{R}^N$ with the $M$ measurement functions $\bm{\phi}_{i}(t)$ for $i=1,\ 2, \cdots,\ M$ evaluated over a finite duration. That is, $y_{i} = \langle \bm{S}(t), \bm{\phi}_{i}(t)\rangle$; for $i=1,\ 2, \cdots,\ M$. The measurement vector is
\begin{equation}
\bm{y} = \bm{\Phi} \bm{\Psi} \bm{C} = \bm{A}\bm{C}= \bm{\Phi} \bm{S}(t),  \label{newmeasurement}
\end{equation}
where $\bm{A}=\bm{\Phi} \bm{\Psi}$, with $\bm{\Phi}$ and $\bm{\Psi}$ having sizes $M \times N$ and $N \times N$, respectively.
To achieve  stable recovery of the sparse set of coefficients and reconstruction of the signal, the measurement functions should be chosen such that they capture maximum information of $\bm{S}(t)$. In general, $K \le M$ and $M \ll N$. For convenience, in the remaining part of the article, we use $\bm{S}$ in place of $\bm{S}(t)$. 
\par To minimize the number of measurements $M$, the measurement functions $\{\bm{\phi}_{i}(t)\}_{_{i=1}}^{^M}$ should not be able to sparsely represent the representation basis $\{\bm{\psi}_j(t)\}_{_{j=1}}^{^N}$ relative to which the signal is sparse, and vice versa \cite{bara}.  Thus, the mutual coherence between the two matrices $\bm{\Psi}$ and $\bm{\Phi}$ denoted by 
\begin {equation}
\mu(\bm{\Phi},\bm{\Psi}) = \sqrt{N}\; max\{\left|\langle \bm{\psi}_j, \bm{\phi}_i \rangle \right|, \; for \;1 \leq i \leq  M, \ 1 \leq j \leq  N \} 
\end {equation}
should be minimum. For normalized matrices, $\mu$ is within $[1,\sqrt{N}]$ \cite{CSC}.
Thus, for compressed sensing of signals, a representation basis in which signal is exactly sparse or is compressible is to be identified. A measurement matrix, which is incoherent with the representation basis, and algorithms necessary to recover the sparse representation coefficients from the compressed measurements constitute the other vital components in CS.   
\par In \cite{Peyre}, Peyre proposes a method to reconstruct the signal from $\bm{y}$ when information about the measurement matrix alone is known. In this method, estimation of representation basis and recovery of sparse signal are done simultaneously. The representation basis which results in maximum sparsity of the signal is estimated from a tree structured dictionary of orthogonal bases using iterative thresholding algorithm. In  \cite{Ravi}, Ravishankar and Bresler discuss a method for learning sparsifying transform from the data. They propose a generalized formulation of transform learning at the analysis side that learns well-conditioned transforms under both noiseless and noisy conditions.
\par The commonly used iterative recovery algorithms assume knowledge of the representation basis, in which the signal is sparse. These  algorithms find an approximation to the signal by minimizing the residual energy 
 \cite{candtao}\cite{mallat}-\cite{stomp} under high signal to noise ratio. Greedy pursuit algorithms like Orthogonal Matching Pursuit (OMP) \cite{ompref}, its variants generalized OMP (gOMP) \cite{omp}, Regularized OMP (ROMP) \cite{romp} and Compressed Sampling Matching Pursuit (CoSaMP) \cite{cosamp} exhibit good performance when sparsity is known in advance. These algorithms use sparsity as a parameter. They also consider unrecoverable energy to be greater than noise and therefore require noise level to be significantly low compared to signal level. In \cite{orthonoise}, compressed sensing of a signal of interest corrupted by an interfering signal is filtered to separate the signal of interest from noise. But orthogonality condition is imposed on the noise subspace with respect to the signal subspace for achieving the desired result. In practical scenario, it is not easy to meet these constraints. These constraints can be removed if we resort to methods which do not consider $\ell_2$ norm directly for choosing the support.
In \cite{bestbasis}, best representation basis is identified adaptively, from a dictionary of wavelet packets by choosing the decomposition structure which minimizes Shannon entropy. 
  \par The entropy minimization based matching pursuit (EMP) algorithm proposed in this article is motivated by the fact that sparsity can be induced by minimizing Shannon entropy of signal representation.  In the sequel, entropy means Shannon entropy. The advantage offered by EMP algorithm is its noise resilience during signal recovery. In the absence of noise, the performance of EMP algorithm is at par with the Matching Pursuit (MP) algorithm in terms of Signal to Reconstruction Error Ratio (SRER) \cite{ourpaper}. The EMP algorithm can be used to arrive at a sparse representation of a signal when its representation basis is known. Sparse representation and signal recovery are considered dual to each other in \cite{ompref}. 
  \par In this article, EMP algorithm is used for signal recovery from measurements in the context of compressed sensing. 
Compared to the conventional greedy pursuit algorithms, the EMP algorithm has superior capability to extract signal components from noisy measurements. In Section 2, we focus on the formulation of the EMP algorithm along with its performance analysis and proof of convergence. 
 Section 3 presents results of simulation study carried out on synthetic sparse signals and a class of signals for which sparsity is not known upfront. The class of signals chosen is speech signals. Results are presented for both noise-free and noisy cases. A discussion on the results is also presented. The article is concluded in Section 4.  
\section{ Entropy minimization based Matching Pursuit Algorithm for signal recovery}
Matching pursuit (MP) algorithm, OMP algorithm, ROMP algorithm and CoSaMP algorithm are examples of greedy iterative pursuit algorithms. Each iteration updates $\bm{\hat{y}}$, the approximation of $\bm{y}$ in (\ref{newmeasurement}), such that the residual energy $\|\bm{y-\hat{y}}\|_2^2$ is minimized. The update $\bm{\hat{y}}$ is obtained by choosing one or more columns from the matrix $\bm{A}$, that correlated best with the residual error vector resulting from the previous iteration. 
\par  The EMP algorithm is a variant of MP Algorithm. It minimizes the overall entropy of the signal representation in each iteration instead of minimizing the residual energy. Entropy $H(\bm{S})$ of the representation of the signal $\bm{S}$ is related to the theoretical dimension $N$ of the signal as \cite{bestbasis, Lubbe}
  \begin{equation} 
N=exp(H(\bm{S})). \label{dim1} 
\end{equation} 
  In \cite{ourpaper}, the EMP algorithm is used for obtaining a sparse representation of a class of signals from its representation in time domain assuming that the sparsifying frame is known. In the context of compressed sensing, we extend the algorithm for recovering the sparse representation $\bm{C}$ of $\bm{S}$ from the measurements $\bm{y}$. 
The estimate $\bm{\hat{C}}$ of $\bm{C}$, thus obtained, is used to obtain an estimate $\bm{\hat{S}}$ of the signal $\bm{S}$ through $\bm{\hat{S}} = \bm{\Psi} \bm{\hat{C}}$.   
   \par Without loss of generality, we consider a normalized signal $\bm{X} = \{x_i, i=1,2,...N\}$. $x_i^2$ represents the probability of choosing the $i$-th function of some basis in $\mathbb{R}^N$. The entropy of representation of $\bm{X}$ is defined by
\begin {equation}
 H(\bm{X})= \sum_{i=1}^N{x_i^2 log \frac{1}{x_i^2}} .
 \label{entr_c}
    \end{equation}
    
Let the entropy of the representation of the signal $\bm{y}$ be denoted as $H(\bm{y})$. At iteration $m$, the conditional entropy of the representation of $\bm{y}$, given the vectors in the set $\bm{\hat{A}}^{(m)}$, is represented as $H(\bm{y|\hat{A}}^{(m)})$, where $\bm{\hat{A}}^{(m)}$ is the matrix which approximates $\bm{A}$ at the $m^{th}$ iteration. Similarly, let the conditional entropy of representation of the residual signal $\bm{e}$, given $\bm{y}$ and $\bm{\hat{y}}$, be $H(\{\bm{y-\hat{y}}\}|\bm{\hat{A}}^{(m)})$ and the conditional entropy of the representation of $\bm{\hat{y}}$, given the vectors in the set $\bm{\hat{A}}^{(m)}$, be $H(\bm{\hat{y}}|\bm{\hat{A}}^{(m)})$. The mutual information between $\bm{\hat{y}}$ and $\bm{\hat{A}^{(m)}}$ is denoted as $I(\bm{\hat{y},\hat{A}}^{(m)})$.
In every iteration, signal $\bm{y}$ is represented as the sum of approximation $\bm{\hat{y}}$ and residual $\bm{e}$ {with the available $\bm{A}$. Aim is to choose the smallest subset of $\bm{A}$  to estimate $\bm{\hat{y}}$.             
\begin{equation}
       \bm{y =\hat{y}+e} \\
       \label{y_rep}
       \end{equation}
       Though $\bm{e}$ is completely determined by $\bm{y}$ and $\bm{\hat{y}}$, sparsity of $\bm{e}$ or representation entropy of $\bm{e}$ will be determined by the choice of $\bm{\hat{A}}^{(m)}$ which determines $\bm{\hat{y}}$.  
      \begin{eqnarray}
     H(\bm{y|\hat{A}}^{(m)}) &=& H(\bm{\hat{y}|\hat{A}}^{(m)}) +  H(\{\bm{y-\hat{y}}\}|\bm{\hat{A}}^{(m)}) 
      \label{eq:entropy2} 
   \end{eqnarray}  
Using the definition of mutual information,
\begin{eqnarray} 
       H(\bm{\hat{y}|\hat{A}}^{(m)}) &=&  H(\bm{\hat{y}})-I(\bm{\hat{y}},\bm{\hat{A}}^{(m)})
\end{eqnarray} 
Since $\bm{\hat{y}=\hat{A}}^{(m)}\bm{\hat{C}}$, $H(\bm{\hat{C}})$ is the information left in $\bm{\hat{y}}$ when prior information about $\bm{\hat{A}}^{(m)}$ is available. Thus,
\begin{eqnarray}    
       H(\bm{\hat{y}|\hat{A}}^{(m)})&=&  H(\bm{\hat{C}})
\label{sparse}
\end{eqnarray}
Substituting (\ref{sparse}) in (\ref{eq:entropy2}), 
     \begin{eqnarray}
      H(\bm{y|\hat{A}}^{(m)}) &=& H(\{\bm{y-\hat{y}}\}|\bm{\hat{A}}^{(m)})+ H(\hat{\bm{C}}).  
      \label{eq:entropy1} 
   \end{eqnarray} 
$H(\{\bm{y-\hat{y}}\}|\bm{\hat{A}}^{(m)})$ is the entropy of representation of residual error. A sparse $\bm{\hat{C}}$ results by minimizing the conditional entropy $H(\bm{\hat{y}|\hat{A}}^{(m)})$.}
 On convergence of the algorithm, $H(\{\bm{y-\hat{y}}\}|\bm{\hat{A}}^{(m)})$ is zero or is negligible compared to $H(\bm{\hat{C}})$.  Thus, minimization of $H(\bm{y|\hat{A}}^{(m)})$ leads to minimum $H(\bm{\hat{C}})$. By (\ref{dim1}), our aim is to minimize $H(\bm{S})$ in order to reduce the dimension of $\bm{S}$ which is achieved by minimizing $H(\bm{y|A})$. This is attained by solving 
\begin {equation}
\min_{\bm{C}} \; \{\textit{H}(\bm{C})\}\;  \textit{subject to}\; {\bm{y}}= \bm{A}\bm{C}; \bm{A=\Phi\Psi}. 
\label{eq:recoveryaim}
\end{equation}
   \par  Since the signal can be normalized, $ 0\le e_i^2\le 1$, where $e_i$ is the $i^{th}$ component of $\bm{e}$, and $e_i^2$ can be considered as the probability of occurrence of the component $e_i$. Thus entropy of the representation of residue conditioned on the estimated $\bm{\hat{y}}$ is calculated as $\sum_{i=1}^M{e_i^2 log \frac{1}{e_i^2}}$. Similarly, $H(\bm{\hat{C}}$) is calculated using the normalized vector of sparse measurements as $\sum_{i=1}^N{\hat{c}_i^2 log \frac{1}{\hat{c}_i^2}}$, where $\hat{c}_i$ is the $i^{th}$ component of $\bm{\hat{C}}$. 
\subsection{ EMP Algorithm for noiseless input signals}
  We have the measurement vector $\bm{y}$, the matrix of representation basis $\bm{\Psi}$ and the matrix of measurement functions $\bm{\Phi}$. They are related by (\ref{newmeasurement}). Our aim as in (\ref{eq:recoveryaim}) is 
 achieved by 
\begin{equation} 
\min_{\bm{\hat{_{C}}}}\; \{H(\bm{y|A})\}\hspace{0.05in}  \textit{subject to} \left\|\bm{y-\hat{y}}\right\|_{2} < \varepsilon, \textnormal{where } \bm{\hat{y}= A} \bm{\hat{C}}; \varepsilon > 0. 
\label{eq:noiselessaim}
\end{equation}
In each iteration of the algorithm we extract one component of the signal, that carries maximum information, from the residual $\bm{e}$ and refine the approximation $\bm{\hat{y}}$.  Only one $\hat{c_i}$ changes in the calculation of $\bm{\hat{y}}$, but all the $M$ components of $\bm{e}$ have the flexibility to change.  Hence, more importance has to be given in reducing $H(\{\bm{y-\hat{y}}\}|\bm{\hat{A}}^{(m)})$ compared to the increase in $H(\bm{\hat{y}|\hat{A}}^{(m)})$. 
    \par Our aim is to capture maximum information of the signal from the residue $\bm{e}$ using each coefficient $\hat{c}_i$ and minimize $H(\{\bm{y-\hat{y}}\}|\bm{\hat{A}}^{(m)})$, thus minimizing the conditional entropy $H(\bm{y|\hat{A}}^{(m)})$. In the absence of noise, the residual $\bm{e}$ contains contributions solely from $\bm{y}$ and hence $\bm{e}$ approaches zero with every iteration. To facilitate the convergence of the algorithm, residual signal energy should decrease in every iteration.\\
\emph {\textbf{EMP Algorithm for noiseless case}}\\
 \framebox{
 \parbox [h]{5.3 in}{
 \textbf{Task}   
   \\      
     To find a sparse representation $\bm{C}$ of a signal $\bm{S}$ in $\bm{\Psi}$ domain
      subject to $\bm{y}= \bm{A}\bm{C},\; \left\|\bm{y-\hat{y}}\right\|_2 < \varepsilon $.\
      $\bm{A}=\bm{\Phi} \bm{\Psi}$, where $\bm{\Phi}$ is the measurement matrix, and $\bm{\hat{y}}= \bm{A} \bm{\hat{C}}$ \
      with $\bm{\hat{C}}$ as the estimated coefficient vector.
     \\
     \textbf{Parameters}
     \\ 
       Given $\bm{A}$ whose columns form a frame, the measurements $\bm{y}$ and error threshold $\varepsilon$. 
       \\
      \textbf{Initialization} 
    \\
    a. Save the norm of $\bm{y}$. \\
    b. Approximation basis set $\bm{\hat{A}}$ with an $ M \times N$ matrix of zeros.
    \\
    c. Measurement vector approximation $\bm{\hat{y}}^{(0)}$ with $M \times 1$ vector of zeros. \\
    d. Estimated representation $\bm{\hat{C}}^{(0)}$ to $N \times 1$ vector of zeros. \\
    e. Iteration index $m$ to 0. \\
    f. Residual vector $\bm{r}^{(0)}, \bm{e}^{(0)}$ to normalized $\bm{y}$. 
\\
    g.	Magnitude of error vector to unity  $\left \|\bm{r}^{(0)}\right \|_{2} =1 $. 
\\
    h.	Set $\varepsilon$ to the given threshold, $w_1=\frac{N}{N+1}$ and $w_2= \frac{1}{N+1}$.
}}
\framebox{
 \parbox [t]{5.3 in}{
\textbf{Main iteration} \\
 Increment $m$ by 1 and perform the following steps. \\
for each column index $j = 1 \;to\; N$ \\
   \hspace{0.1in}\{ \\
   \hspace{0.3in} $c$ = $\left\langle \bm{r}^{(m-1)},\bm{A_j}\right\rangle$, $\bm{A_j}$ is the $j^{th}$ column of $\bm{A}$ \\
   \hspace{0.3in}  $\bm{\hat{y}}^{(m)}_{temp} = \bm{\hat{y}}^{(m-1)}+c\bm{A_j}$ \\
   \hspace{0.3in}  $\bm{e= y-\hat{y}}^{(m)}_{temp}$ \\
   \hspace{0.3in}  Calculate $H(\bm{e}|\bm{\hat{A}}^{(m)})$ = $\sum_i{e_i^2 log \frac{1}{e_i^2}}$  and \\
   \hspace{0.35in}  $H(\bm{\hat{y}}|\bm{\hat{A}}^{(m)})$ = $\sum_i{\hat{c}_i^2 log \frac{1}{\hat{c}_i^2}}+ c^2 log \frac{1}{c^2}$ for normalized coefficients\\
   \hspace{0.3in} Find the index $j_0$ which minimizes \\
   $H(\bm{y|\hat{A}}^{(m)})= (w_1 H(\bm{e}|\bm{\hat{A}}^{(m)}) + w_2H(\bm{\hat{C}})) $ and $\left \|\bm{e}^{(m)} \right \|_2$ $<$ $ \left \|\bm{e}^{(m-1)} \right \|_2$.   \\ 
   \hspace{0.1in}\} \\
   \textbf{Update Support, residual signal and representation vector} \\
   $\bm{\hat{A}}^{(m)} = \bm{\hat{A}}^{(m-1)}$ replaced with $\bm{A}_{j_0}$ at the $j_0^{th}$ position.\\
     $\bm{e}^{(m)} = \bm{e}^{(m-1)} - \langle \bm{e}^{(m-1)}, \bm{A}_{j_0} \rangle \bm{A}_{j_0}$ \\
     $\hat{c_{j_0}} = \langle \bm{e}^{(m-1)}, \bm{A}_{j_0} \rangle $ \\
     $\bm{\hat{C}}^{(m)} = \bm{\hat{C}}^{(m-1)}$ added with $\hat{c_{j_0}}$  at the $j_0^{th}$ position.      \\
     $\left \|\bm{e}^{(m)} \right \|_{2} = \left(\sum_{i=1}^N |\bm{e}^{(m)}_i |^2\right)^{\frac{1}{2}} $\\
     $\bm{r}^{(m)}=\bm{e}^{(m)}$
 \\
  \textbf{Stopping rule } \\
     \hspace{0.15in} If $\left \|\bm{e}^{(m)} \right \|_{2} < \varepsilon $, stop. Otherwise apply another iteration. \\
     \\
     \textbf{Output}\\
      \hspace{0.15in} Required sparse representation in $\bm{\hat{C}}$.  Norm of the signal is restored using the value saved before normalization step. Reconstructed signal $\bm{\hat{S}}= \bm{\Psi\hat{C}}$.
    }}
\par Entropy reduction need not always minimize the residual signal energy in each iteration. An iteration could choose a vector $\bm{A}_i$ from $\bm{A}$, orthogonal to the error vector, which will not reduce the error energy.  But this situation is avoided in this algorithm by rejecting vectors from $\bm{A}$ which do not reduce the residual energy. To this end, the residual signal is projected onto the chosen vector $\bm{A}_i$, and a new component is added to the approximated signal only if the component along the direction of $\bm{A}_i$ reduces the residual energy. Otherwise $\bm{A}_i$ is discarded. A vector $\bm{A}_i$ which minimizes the conditional entropy $H(\bm{y|\hat{A}}^{(m)})$ as in (\ref{eq:entropy1}) with reduced error is selected in each iteration. Iterations continue till $\left \|\bm{y-\hat{y}}\right \|_2 < \varepsilon$, where $\varepsilon$, the error tolerance permitted by the application, has a value close to zero.
     \subsection{Convergence of EMP Algorithm}
  We consider a signal $\bm{S}$ with its representation $\bm{C} \in \mathbb{R}^N$. A finite dimensional vector space is complete. Since $\bm{\Psi}$ is a normalized basis in $\mathbb{R}^N$, $\bm{S}$ can be represented as a linear combination of the elements of $\bm{\Psi}$ with zero residual. Convergence of the algorithm is proved by showing that the signal approximation, $\bm{\hat{S}}^{(m)}$, due to the iterations in the algorithm, results in a Cauchy sequence in $\mathbb{R}^N$. Here we assume that $\bm{\Phi}$ satisfies the restricted isometry property for sparsity $2K$, thus guaranteeing unique recovery of a $K$ sparse signal. Further, we assume that the RIC for sparsity $K$ is very small. Hence the problem of finding $\bm{S}$ from $\bm{y}$ reduces to the problem of finding $\bm{C}$ from $\bm{S}$. Selecting a column from the matrix $\bm{A}$ is equivalent to the selection of the corresponding column in $\bm{\Psi}$.
  In the beginning of the first iteration, the residual $\bm{r}^{(0)}$ = $\bm{S}$.  In iteration ($m+1$),
     $\bm{r}^{(m)}= \bm{S}-\bm{\hat{S}}^{(m)}$ can be expressed as
  \begin{eqnarray} 
     \bm{r}^{(m)}=\left \langle \bm{r}^{(m)}, \bm{\psi}_{i(m+1)} \right \rangle \bm{\psi}_{i(m+1)} + \bm{r}^{(m+1)}, 
      \label{eq:conv3}    
   \end{eqnarray} 
where $\bm{\psi}_{i(m+1)}$ is the $i$-th column of $\bm{\Psi}$ selected in the $(m+1)$-th iteration, $\bm{\hat{S}}^{(m)}$ is the approximation of $\bm{S}$ in the $m$-th iteration and $\bm{r}^{(m)}$ is the residual which resulted out of $\bm{\hat{S}}^{(m)}$.
   \begin{eqnarray} 
    \left | \langle \bm{r}^{(m)}, \bm{\psi}_{i(m+1)} \rangle \right |^2 =\left \| \langle \bm{r}^{(m)},\bm{\psi}_{i(m+1)} \rangle \bm{\psi}_{i(m+1)} \right\|_{2}^2, \; \textnormal{since}\; \left \| \bm{\psi}_{i(m+1)} \right \|^2_{2}=1. 
    \label{eq:norm}    
   \end{eqnarray}
  The residual $\bm{r}^{(m+1)}$ is orthogonal to $\bm{\psi}_{i(m+1)}$. Using (\ref{eq:conv3}) and (\ref{eq:norm}),   
    \begin{eqnarray} 
     \left \|\bm{r}^{(m)}\right \|_{2} ^2 = \left| \langle \bm{r}^{(m)}, \bm{\psi}_{i(m+1)} \rangle \right|^2 +\left \| \bm{r}^{(m+1)} \right \|^2_{2}. 
      \label{eq:conv4}    
   \end{eqnarray}  
Using (\ref{eq:conv3}) to (\ref{eq:conv4}), 
  \begin{eqnarray} 
     \left \| \bm{S} \right \|^2_{2} =\left \|\bm{r}^{(0)} \right \|^2_2 = \sum^{N}_{m=0}\left| \langle \bm{r}^{(m)},\bm{\psi}_{i(m+1)} \rangle  \right|^2 + \left \| \bm{r}^{(N+1)}\right \|^2_{2}.
      \label{eq:conv5}    
   \end{eqnarray} 
From (\ref{eq:conv4}),
  \begin{eqnarray} 
     \left \| \bm{r}^{(m+1)} \right \|^2_{2}= \left \|\bm{r}^{(m)}\right \|_{2} ^2 - \left| \langle \bm{r}^{(m)}, \bm{\psi}_{i(m+1)} \rangle \right|^2.  
      \label{eq:conv6}    
   \end{eqnarray} 
   Hence,  $\frac{\left \| \bm{r}^{(m+1)}\right \|^2_{2}}{\left \| \bm{r}^{(m)}\right \|^2_{2}} = 1 - \frac{\left| \langle \bm{r}^{(m)}, \bm{\psi}_{i(m+1)} \rangle \right|^2}{\left \|\bm{r}^{(m)}\right \|_{2} ^2}  \le 1$. 
Equality arises when the chosen vector is orthogonal to $\bm{r}^{(m)}$. Orthogonal vectors are discarded in the algorithm as they will not refine the residual signal. 
Thus, \{$\bm{r}^{(m)};\; m=1,2,...$\} is a bounded decreasing sequence making \{$\bm{\hat{S}}^{(m)};\; m=1,2,...$\} a bounded increasing sequence bounded above at $\left \|\bm{S} \right \|_2$. Since $\left \|\bm{\hat{S}}^{(m)} \right \|_2$ is not the upper bound, there exists an integer $L$ such that
\begin{equation}  
\left \|\bm{S}-\bm{\hat{S}}^{(m)}\right \|_2 \leq \left \|\bm{S}-\bm{\hat{S}}^{(L)}\right \|_2,\; \textnormal{for}\; m \geq L. 
\end{equation}
Thus,
\begin{equation} 
\left \|\bm{S}-\bm{\hat{S}}^{(m+1)}\right \|_2 \leq \left \|\bm{S}-\bm{\hat{S}}^{(m)}\right \|_2 \leq \left  \|\bm{S}-\bm{\hat{S}}^{(L)}\right \|_2,\; \textnormal{for}\; m \geq L.
\end{equation} 
Hence \{$\bm{\hat{S}}^{(m)},\; m=1,2,...$\} is Cauchy. Equivalently, \{$ \|\bm{\Psi}(\bm{C}-\bm{\hat{C}}^{(m)})  \|_2; \; m=1,2,...$\} is a bounded decreasing sequence which implies that the set of representation coefficients, \{$\bm{\hat{C}}^{(m)};\; m=1,2,...$\}, forms a Cauchy sequence.
\par The EMP algorithm may not give the exact representation with the sparsest set of coefficients in a finite number of iterations unless the vectors are orthonormal. In the noise-free case, it is possible to recover the sparse set of coefficients ideally, as convergence of the algorithm is guaranteed. The algorithm can be modified to calculate the vector of sparse representation coefficients, $\bm{\hat{C}}$, in a finite number of iterations. This calculation is based on the minimization of $\left \|\bm{y-\hat{y}} \right\|^2$, after choosing a candidate vector from $\bm{A}$ by minimizing $H(\bm{y|\hat{A}}^{(m)})$, in each iteration. This can lead to recovery in $K$ iterations, where $K$ is the sparsity of the signal, as guaranteed by the OMP algorithm \cite{ompref}. 
  \subsection{ EMP Algorithm for noisy input signals}
  When the input signal is noisy, the aim of EMP algorithm in each iteration is to reduce the overall conditional entropy 
  calculated as in (\ref{eq:entropy1}). We assume white noise having dense representation $ \tilde{\bm{C}}$ relative to the basis $\bm{\Psi}$ in which the signal has a sparse representation $\bm{C}$. The noisy signal $\bm{S}$ is
\begin {equation}
   \bm{S}= \bm{S_o +S_n}, 
\end {equation}
where $\bm{S_o}$ and $\bm{S_n}$ are the signal and noise components, respectively. The corresponding noisy measurement vector is
\begin{equation}
   \bm{y} = \bm{y_o + y_n},
\end{equation}
where $\bm{y}_o$ and $\bm{y}_n$ are the signal and noise components, respectively, in the measurement vector. Using the notions of representation and measurement,
\begin{equation}
    \bm{y_o}= \bm{\Phi S_o} = \bm{\Phi} \bm{\Psi} \bm{C} \\
\end{equation}
\begin{equation}    
    \bm{y_n} = \bm{\Phi S_n} = \bm{\Phi} \bm{\Psi} \tilde{\bm{C}} \\
\end{equation}    
\begin{equation}
    \bm{y}= \bm{\Phi} \bm{\Psi} (\bm{C}+\tilde{\bm{C}}) =\bm{A}(\bm{C}+\tilde{\bm{C}}), 
\end{equation}
where $\bm{A}=\bm{\Phi} \bm{\Psi}$ is of full row rank, and hence its pseudo-inverse exists. The vector $\bm{C}$ is sparse but $\tilde{\bm{C}}$ is dense as the representation basis represents signal sparsely and noise densely. If $\mathcal{R}$ represents a recovery algorithm, then $\mathcal{R}(\bm{y}) =\bm{C}$. Conventional recovery algorithms work on least square error minimization and hence are not capable of distinguishing between $\bm{C}$ and $\tilde{\bm{C}}$. The only distinguishing factor between $\bm{C}$ and $\tilde{\bm{C}}$ is that $\bm{C}$ is sparse and $\tilde{\bm{C}}$ is dense relative to the chosen representation basis $\bm{\Psi}$. EMP algorithm makes use of entropy minimization method which has the inherent capability to distinguish and extract sparse coefficients. The component $\tilde{\bm{C}}$ in the sum $\bm{C}+\tilde{\bm{C}}$ increases the entropy of representation. In the proposed EMP algorithm, noise components are rejected such that the increase in conditional entropy of signal representation is restricted below a predefined limit $\gamma$, in each iteration of the algorithm. 
{
  When $\bm{C}$ and $\tilde{\bm{C}}$ are normalized,}
    \begin {equation}
    H(\bm{C}) < H(\tilde{\bm{C}}),
    \label{sparse_c}
\end{equation}
where $H(\bm{C})$ and $H(\tilde{\bm{C}})$ are respectively, the entropy of $\bm{C}$ and $\tilde{\bm{C}}$  calculated according to (\ref{entr_c}).   
The morphological component separation presented in \cite{MCA1} can be used to separate noise and signal components based on their respective sparsity in a given basis. 
The goal of the EMP algorithm under noisy conditions is
\begin{equation} 
  \min_{\bm{\hat{_{C}}}}\; {H(\bm{y|A})}\hspace{0.05in}  \textit{subject to} \left\|\bm{y-\hat{y}}\right\|_{2} < \varepsilon, and \;\Delta H(\bm{y}) <\gamma, 
   \label{eq:noisyaim}
\end{equation}
\noindent \textnormal{where }$\bm{\hat{y}}= \bm{A} \bm{\hat{C}},\; \Delta H(\bm{y}) = H(\bm{y|\hat{A}}^{(m)}) / H(\bm{y|\hat{A}}^{(m-1)}), \;\gamma > 0,\; \varepsilon > 0.  $ 
  \par The algorithm starts with the measured signal $\bm{y}$ as residual $\bm{e}$, having a maximum of $M$ nonzero elements, and the estimated coefficients $\bm{\hat{C}}$ as a zero vector having $N$ elements. The measurement $\bm{y}$ which is the sum of its approximation and residual, is thus represented using $M$ coefficients out of the overall $N+M$ coefficients. Initial iterations result in significant non-zero coefficients in $\bm{\hat{C}}$, with each nonzero coefficient capturing the $\bm{y}_o$ component from $\bm{y}$. In iteration $m$, 
  \begin{equation}
  \bm{y = \hat{y}}^{(m)} + \bm{e}^{(m)},  \\
 \end{equation}
\begin{flushleft}
where
\end{flushleft}
\begin{equation}
  \bm{\hat{y}}^{(m)}= \sum_i^m{\bm{A}_i \bm{\hat{C}_i}}.
  \end{equation}
  
 The entropy $H(\bm{\hat{C}})$ increases in every iteration as a non-zero component gets added to $\bm{\hat{C}}$. Since EMP algorithm works on minimizing conditional entropy, the column of matrix $\bm{A}$ that induces sparsity in the resulting residue is chosen in each iteration. The conditional entropy $H(\{\bm{y-\hat{y}}\}|\bm{\hat{A}}^{(m)})$ of error, given the measurement vector and its approximation,  decreases as components of signal present in the residue are transferred to $\bm{\hat{y}}$ and captured in $\bm{\hat{C}}$. The energy of the residual decreases considerably. Initial iterations result in overall decrease or at the most marginal increase in $H(\bm{y|\hat{A}}^{(m)})$. Removal of the $\bm{S_o}$ components from $\bm{S}$ causes removal of information about the signal from the residue, thus minimizing its conditional entropy $H(\bm{y|\hat{A}}^{(m)})$. Conditional entropy of $\bm{y}$ is reduced significantly by representing its $\bm{y}_o$ component using sparse $\bm{\hat{C}}$. Therefore, initial iterations capture $\bm{S_o}$ in approximating $\bm{y}$ by $\bm{\hat{y}}$. 
\par The components of noise along the basis chosen in each iteration are also present in the representation. Subsequent iterations have significant noise in the residue $\bm{y}_n$ due to $\bm{S_n}$ which cannot be sparsely represented by $\bm{\hat{C}}$, relative to $\bm{\Psi}$. Therefore, reduction in the first term in (\ref{eq:entropy1}) is considerably lower than the increase in the second term which effectively leads to  increase in $H(\bm{y|\hat{A}}^{(m)})$. In the absence of noise, the energy of the residual signal would not exceed $\varepsilon$, and  $H(\bm{y|\hat{A}}^{(m)})\approx H(\bm{\hat{y}}|\bm{\hat{A}}^{(m)})$ when the algorithm converges. 
 \par When the input is noisy, prior to choosing a particular vector for refining the representation, an additional step in the algorithm compares $H(\bm{y|\hat{A}}^{(m)})$ and $H(\bm{y|\hat{A}}^{(m-1)})$ which are the conditional entropy of the measurement given the chosen subset of vectors in $\bm{\hat{A}}$, in iteration $m$ and $(m-1)$, respectively. The algorithm proceeds to update only if  \\
 $\Delta H(\bm{y})= H(\bm{y|\hat{A}}^{(m)})/H(\bm{y|\hat{A}}^{(m-1)})$ is less than $\gamma$.  The parameter $\gamma$ decides whether a new component is to be added to the signal approximation at the cost of increase in the conditional entropy. It increases linearly with the SNR of input signal allowing more components to be added to the representation. In the noiseless case, $\gamma$ is infinite which reduces residual energy even if it causes increase in the conditional entropy between successive iterations as in (\ref{eq:noiselessaim}). In the noisy case, components which cause increase in the conditional entropy beyond $\gamma$ is attributed to noise components and are rejected by the algorithm. Iterations terminate when either the increment in the conditional entropy between consecutive iterations is beyond the permitted limit $\gamma$ or, the error energy threshold requirement is met.\\
  \\
 \emph {\textbf{EMP Algorithm for noisy case}}\\
 \framebox{
 \parbox [t]{5.3 in}{
 \textbf{Initialization}\\
 Set $w_1=\frac{M-\|\bm{\hat{C}}\|_0}{M}$ and $w_2= \frac{\|\bm{\hat{C}}\|_0}{M}$.\\
 \hspace{2in} \textit{(Other steps are the same as in the algorithm for the case of noiseless measurements)} \\
  \textbf{Main iteration} \\
for each column index $j = 1 \;to\; N$ \\
   \hspace{0.1in}\{ \\
 \hspace{2in}  \textit{(steps are the same as in the algorithm for the case of noiseless measurements)}
     \\    \hspace{0.1in}\} \\
   $\Delta H(\bm{y}) = \frac{H(\bm{y|\hat{A}}^{(m)})}{H(\bm{y|\hat{A}}^{(m-1)})}$. \\
   Stop the iterations if $\Delta H(\bm{y}) < \gamma$, else proceed to next section.\\
   \textbf{Update Support, residual signal and representation vector} \\
\hspace{0.6in}  \textit{(remaining steps are the same as in the algorithm for the case of noiseless measurements)}
  }}
  \\
\section {Results and Discussion}
 The scheme for compressed sensing which uses the EMP algorithm for recovery is studied using both synthetic test signals and actual speech signals. Speech signal is used as an example of a compressible signal. Performance of the scheme under both noisy and noiseless conditions is evaluated. Recovery percentage is used for performance evaluation under noise-free conditions when sparsity is known. Signal to Reconstruction Error Ratio (SRER) is used as an objective measure of performance of the scheme under noiseless condition when sparsity is not known. {It gives a measure of capability of the algorithm to approximate the input signal $\bm{S}$ from the measurements $\bm{y}$ faithfully.} Signal to Noise Ratio (SNR) is used for evaluating the performance under noisy environment. {SNR measures the ability of the algorithm to reject noise in the input signal $\bm{S}$. When the algorithm has ability to reject noise, reconstruction from measurements from a noisy signal $\bm{S}$ will result in low SRER and high SNR.} Since the EMP algorithm aims to minimize conditional entropy and not energy of the residual, we use information power as another parameter for performance comparison.
  Information power (IP) is defined as the variance of a Gaussian source required
to make its entropy equal to that of the entropy $H(\bm{X})$ of the signal source $\bm{X}$. \textit{i.e;} $H(\bm{G})=H(\bm{X})$, where $H(\bm{G})$ is the entropy of the Gaussian source. Reduction in IP indicates sparse or compressible representation. SRER is calculated as 
\begin{equation}
SRER = 10\log_{10} \frac{\sum_i{S_i^2}}{\sum_i{(S_i-\hat{S_i})^2}}, \\
\end{equation}
where $S_i$ represents samples of input signal and $\hat{S_i}$ represents samples reconstructed from the recovered sparse set of coefficients. SNR is calculated as 
\begin{equation}
SNR = 10\log_{10} \frac{\sum_i{S_i^2}}{\sum_i{(S_i-\hat{\tilde{S_i}})^2}}, \\
\end{equation}
where $S_i$ represents samples of input signal before adding noise and $\hat{\tilde{S_i}}$ represents samples reconstructed from the sparse estimate of the representation coefficients based on measurements from noisy signal. The entropy of a Gaussian source with variance $\sigma$ is $H(\bm{G}) = \log_b{\sqrt{2\pi e \sigma^2}}$, where $b$ is the base to be considered.
 Information power is  
 \begin{equation}
 \sigma^2 = \frac{2^{2H(\bm{X})\log_2{b}}}{2\pi e}.
 \label{ip}
 \end{equation}
  Simulations were done with noisy signals generated by adding white Gaussian noise with input SNR varying from $-6dB$ to $3dB$. 
  \subsection{Experiment with synthetic signal}
\begin{figure}[p]
\centering
\includegraphics[width=9.5cm]{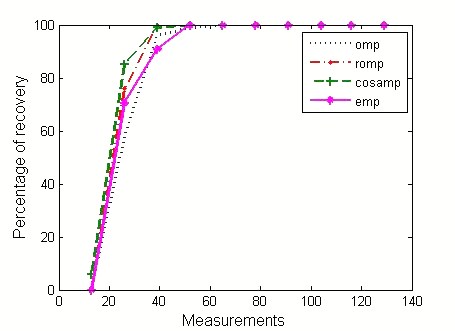}
\caption{Performance comparison of various recovery algorithms on a 4-sparse signal of dimension 200 under noiseless conditions.}
\label{comp_nonoise}
\end{figure}

 Simulations were carried out to study the performance of various recovery algorithms under the following cases.
 \begin{enumerate}
	\item Sparse input signal with orthogonal representation basis and known sparsity.
	\item Sparse input signal with orthogonal representation basis and unknown sparsity.
	\item Sparse input signal with non-orthogonal representation basis added with $3dB$ white Gaussian noise and unknown sparsity.
	\item Compressible input signal with non-orthogonal representation basis added with $3dB$ white Gaussian noise.
\end{enumerate}
  Synthetic test signals are generated by linear combination of representation basis vectors. Fourier representation basis was used as orthogonal basis for the study of case-1. 
The sparsity of the signal considered was $K = 4$. 
Fig.~\ref{comp_nonoise} shows that the performance of EMP algorithm is at par with other recovery algorithms in the absence of noise.

\begin{figure}[p]
\centering
\includegraphics[width=9.5cm]{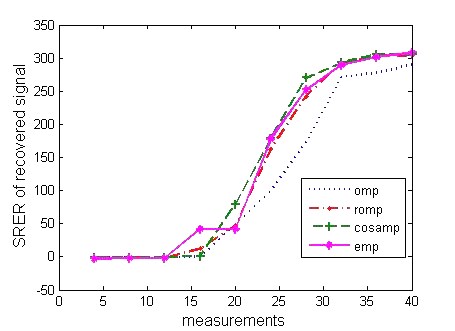}
\caption{Performance comparison of various recovery algorithms on a 4-sparse signal of dimension 200 under noiseless conditions, with unknown sparsity.}
\label{comp_nonoise_spunknown}
\end{figure}

 Fig.~\ref{comp_nonoise_spunknown} shows the performance of various algorithms when sparsity is not known in advance. In this study also, a signal of sparsity $K=4$ was used. But the algorithms were not presented with the information about sparsity. Conventional greedy algorithms estimate sparsity from the measurements. The results show that EMP performs marginally better than the other algorithms when number of measurements is low.
\par The performance of various algorithms in the presence of $3dB$ noise is presented in Table 1. In this study, sparsity of the signal is considered unknown and non-orthogonal representation basis is used. The experiment was carried out for signals having dimension 40 and sparsity $K=4$; the sparsity was not an input to algorithms. The results show that at low SNR, OMP performs better than all other conventional greedy algorithms. EMP outperforms OMP for reduced number of measurements and is marginally better at increased number of measurements. Estimated sparsity was used for halting OMP, CoSaMP and ROMP. The  estimated sparsity varies from 3 to 5 corresponding to measurements varying from 20 to 36. It is observed that as the algorithms approximate the signal with more components beyond its sparsity, more noise gets into the signal approximation thus causing reduction in SNR of the reconstructed signal. Since greedy pursuits estimate sparsity from the number of measurements, increase in the number of measurements leads to increase in the estimated sparsity, resulting in decrease of SNR. This is avoided in EMP as the algorithm does not iterate using sparsity as a parameter. 
\par The performance of various algorithms for a compressible signal in the presence of $3dB$ noise is presented in Table 2.
Since the signal is not strictly sparse, the halting condition of conventional algorithms is changed such that the algorithms terminate when residual norm is below the predetermined threshold. The results indicate that performance of EMP is superior to all other greedy pursuits when the signal is compressible.
 SNR was averaged over  50 runs with different noisy inputs for studying the performance of the algorithms under noisy conditions. 
\begin {table}[ht]
  \centering
  \caption {Comparison of SNR for synthetic sparse signal of length 40 with sparsity unknown, recovered using various greedy pursuits at 3dB noise }
\begin{tabular}{ >{\centering\arraybackslash}p{4cm}  >{\centering\arraybackslash}p{1.8cm}  >{\centering\arraybackslash}p{1.8cm} >{\centering\arraybackslash} p{2cm} >{\centering\arraybackslash} p{1.8cm} }
 \hline
 \centering & \multicolumn{4}{c}{Reconstruction SNR in dB} \\ \cline{2-5}
 \centering{ No. of measurements}& { EMP }  & {OMP} & { CoSaMP}  & { ROMP}  \\
 \hline
  
 20&0.77& 0.02 & -0.59 &0.35 \\
 24&1.74& 0.40 &-0.02 & 1.43\\
 28&2.57& 1.42 & -0.01 & 1.45  \\
 32&3.21& 3.10 & 1.87 & 1.55   \\
 36&3.61& 3.27 & -0.79 & 0.87\\
    
 \hline
  \end{tabular}
   \end {table}

 \begin {table}[ht]
  \centering
  \caption {Comparison of SNR for synthetic compressible signal of length 40, recovered using various greedy pursuits at 3dB noise}
\begin{tabular}{ >{\centering\arraybackslash}p{4cm} >{\centering\arraybackslash} p{1.8cm}  >{\centering\arraybackslash} p{1.8cm} >{\centering\arraybackslash} p{2cm} >{\centering\arraybackslash} p{1.8cm} }
 \hline
 \centering & \multicolumn{4}{c}{Reconstruction SNR in dB} \\ \cline{2-5}
 \centering{ No. of measurements}& { EMP }  & {OMP} & { CoSaMP}  & { ROMP}  \\
  \hline
  
 20&2.33& 2.07 & 1.14 &-0.07  \\
 24&2.65& 2.83 & 2.42 &0.86\\
 28&2.93& 2.70& 1.53 & -0.83  \\
 32&2.88& 2.84 & 1.41 & -0.77  \\
 36&3.21& 3.03 & -0.21 & -0.62\\

 \hline
  \end{tabular}
   \end {table}  
  \begin {table}[ht]
  \centering
  \caption {Comparison of IP for compressible signal of dimension 40 at 0dB noise recovered using various greedy pursuits with 36 measurements.}
\begin{tabular}{>{\centering\arraybackslash}p{4cm} >{\centering\arraybackslash}p{2.5cm} >{\centering\arraybackslash}p{2.5cm} }
 \hline
 {Algorithm}& {IP}  & {SNR in dB } \\
 \hline
  EMP & 3.06 & 3.94 \\
 OMP & 9.53 & 1.87\\
 CoSaMP&61.28&1.10\\
 ROMP &50.32&0.98\\
 \hline
  \end{tabular}
   \end {table}
When the sparsity is unknown, it is estimated from the number of measurements, $M$, and used in the algorithms as $K= M/(2\log_e{N})$, where $N$ is the signal dimension. When non-orthogonal representation basis is considered, the performance of greedy pursuits that update more than one component in an iteration is poor in comparison with the OMP algorithm which chooses only one component per iteration. 
\par As entropy decreases, information power also decreases. Table 3 shows information power of a noisy signal at $0\; dB$ reconstructed using algorithms under consideration. Information power of the original signal without noise was 2.17 and information power of noisy signal was 20.51. The base $b$ considered for logarithm in (\ref{ip}) was 2. Information power  clearly indicates the ability of EMP to reject noise components, that require dense representation and thus have high entropy.

\begin{figure}
\def\tabularxcolumn#1{m{#1}}
\begin{tabularx}{\linewidth}{@{}cXX@{}}
\begin{tabular}{cc}
\subfloat[Input SNR = -6dB]{\includegraphics[width=7cm]{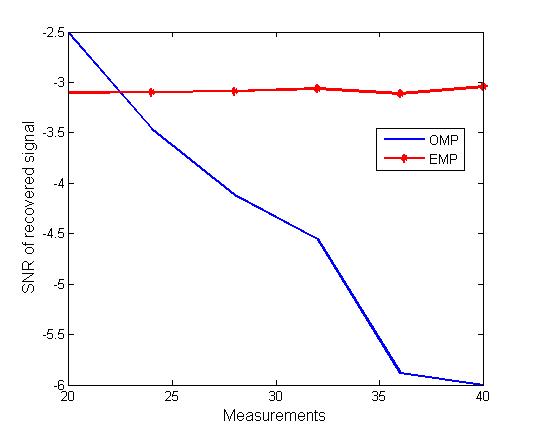}} 
   & \subfloat[Input SNR = -3dB]{\includegraphics[width=7cm]{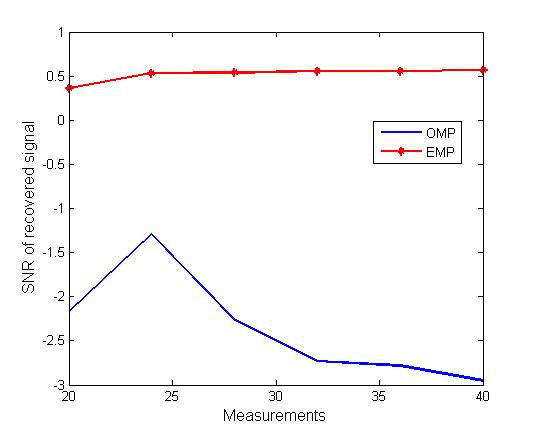}}\\
\subfloat[Input SNR = 0dB]{\includegraphics[width=7cm]{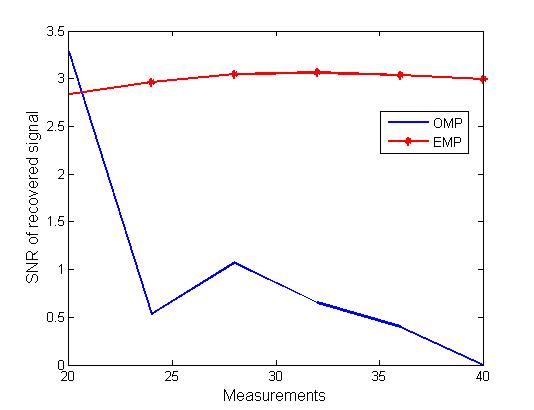}} 
   & \subfloat[Input SNR = 3dB]{\includegraphics[width=7cm]{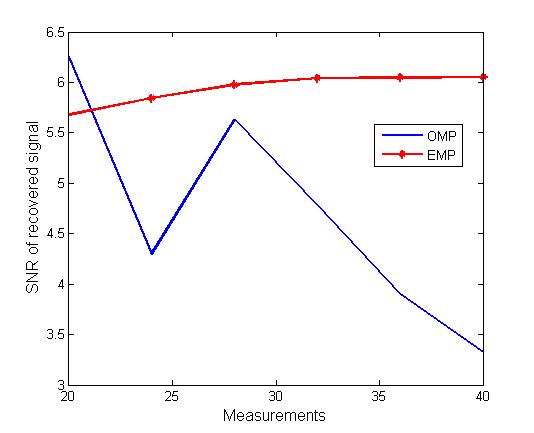}}\\
\end{tabular}
\end{tabularx}
\caption{Performance comparison of EMP and OMP as recovery algorithms on a noisy synthetic signal at various noise levels. }
\label{fig:4}
\end{figure}

  Fig.~\ref{fig:4} shows comparison of the performances of OMP and EMP algorithms for various levels of input noise. Non-orthogonal representation basis was used to generate the synthetic signal. 
Simulation study shows that performance of EMP at SNR ranging from $-6dB$ to $3dB$ is higher than that of all other algorithms considered, indicating the capability of the EMP algorithm for providing robustness to the CS system in the presence of noise.

\subsection{Experiment with speech segment}
This experiment makes use of the representation basis obtained in \cite{ourpaper}.  A speech segment sampled at Nyquist rate is used as input. Wavelet packet basis is used as the representation basis. Wavelet packet basis obtained using the best basis algorithm proposed in \cite{bestbasis} is chosen to arrive at the tree structure suitable for speech signals \cite{ourpaper}. Impulse response at each of the terminal nodes of the tree and their translates constituted the representation basis. The details of the signal representation can be found in \cite{ourpaper}. The representation basis is non-orthogonal. Measurement matrix is chosen based on the Multi-Dimensional Scaling method proposed in \cite{mdsproof}. This experiment benchmarks the performances of the algorithms when the signal is compressible in a non-orthogonal basis with its sparsity unknown. 
Perceptual quality of the reconstructed signal was indistinguishable from the original in the noise-less case for the number of measurements as low as 60\% of that required at Nyquist rate. Table 4 shows that the performance of the EMP algorithm is at par with the OMP algorithm under noise-less case.
  \begin {table}[ht]
  \centering
  \caption {Comparison of SRER (in dB) for speech signal divided into segments of length 40, recovered through the OMP and the EMP algorithms.}
\begin{tabular}{ >{\centering\arraybackslash}p{6cm}  >{\centering\arraybackslash}p{3cm}  >{\centering\arraybackslash}p{3cm} }
 \hline
 {No. of measurements }& {OMP }  & {EMP}  \\
  \hline
 16 & 2.93 & 3.30  \\
 20& 7.77&  8.03\\
 24& 11.32& 11.76\\
 28 & 13.44& 13.77 \\
 32 & 17.96 & 18.03 \\
 38 & 26.83 & 27.30 \\
 40 & 267.74 & 267.60 \\
 \hline
  \end{tabular}
   \end {table}
\par
\begin{figure}
\def\tabularxcolumn#1{m{#1}}
\begin{tabularx}{\linewidth}{@{}cXX@{}}
\begin{tabular}{cc}
\subfloat[Input SNR = -6dB]{\includegraphics[width=7cm]{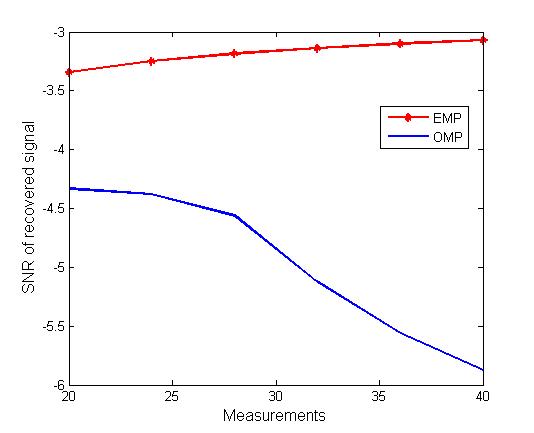}} 
   & \subfloat[Input SNR = -3dB]{\includegraphics[width=7cm]{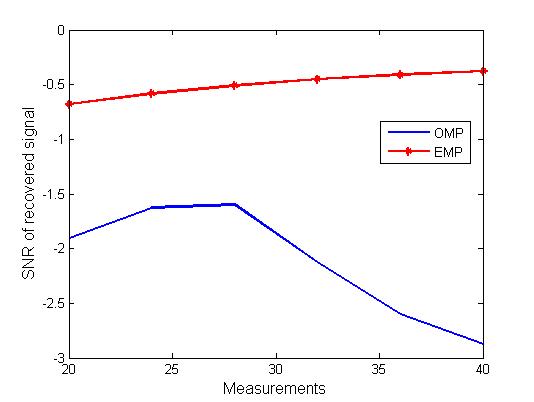}}\\
\subfloat[Input SNR = 0dB ]{\includegraphics[width=7cm]{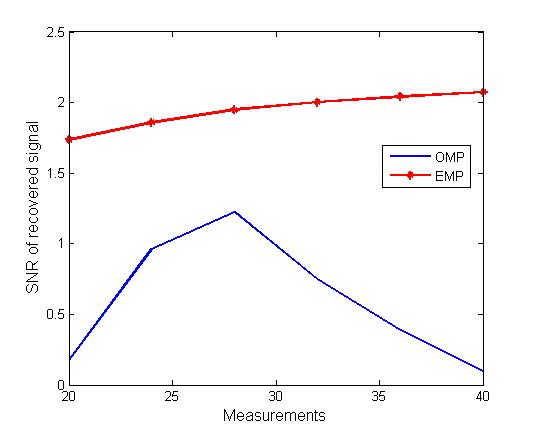}} 
   & \subfloat[Input SNR = 3dB]{\includegraphics[width=7cm]{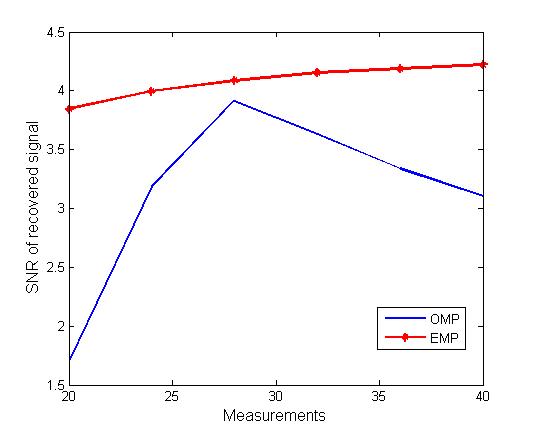}}\\
  
\end{tabular}
\end{tabularx}

\caption{Performance comparison of the EMP and the OMP recovery algorithms on a noisy speech signal at various noise levels. }
\label{fig:5}
\end{figure}

 Fig.~\ref{fig:5} shows the performance when speech signal with noise is compressively measured and reconstructed using the EMP and the OMP algorithms. The halting condition of the OMP algorithm was based on residual error threshold and not based on the estimated sparsity since the input signal was compressible and not exactly sparse. The figure shows the superior performance of the EMP algorithm.
 
 The simulation results show that the EMP algorithm  has the capability to perform at par with the CoSaMP, ROMP and OMP algorithms when the signal is sparse and without any noise. But, for noisy compressible signals, the EMP algorithm performs significantly better than the other greedy algorithms mentioned.  With low SNR, reducing number of measurements leads to smaller estimated sparsity which improves the performance of the OMP, CoSaMP and ROMP algorithms. Since the EMP algorithm tries to reduce conditional entropy, it has the ability to reject noise that requires dense coefficients in the chosen representation basis for the signal. The performance is more or less independent of the number of measurements. But computational overhead of the EMP algorithm is higher than that for the other algorithms as entropy calculation is computationally intensive. In the EMP algorithm, the chosen threshold value $\varepsilon$ has to meet the constraints imposed by applications on the required minimum SRER under noiseless case. When noisy signal is presented, the relaxation parameter $\gamma$ should be fine-tuned in accordance with the SNR of the input signal. In our study, $\gamma$ was calculated as $(M+N+5 SNR)/M$, where SNR is in dB.
       
   \section {CONCLUSION } 
In this article, we have presented the EMP algorithm for sparse signal recovery under noiseless and noisy conditions. The sparse recovery scheme based on the proposed EMP algorithm is unique in its robustness in the presence of noise. The classical greedy algorithms reconstruct the original signal faithfully but fail to separate noise. The functionality of the EMP algorithm is based on conditional entropy minimization instead of energy minimization which facilitates its noise resilience. The EMP Algorithm does not require the sparsity to be known, at the same time, it offers significant improvement in the SNR of the reconstructed signal in the presence of noise. Noise components do not get added to the reconstructed signal even when measurements are increased. Therefore, the EMP algorithm may replace the conventional greedy pursuit algorithms when the measurements are noisy. Under noiseless conditions, the performance of the EMP algorithm is at par with the best performance of other greedy algorithms. 

\end{document}